\documentclass[10pt,twocolumn,letterpaper]{article}

\usepackage[pagenumbers]{cvpr}
\usepackage{float}

\definecolor{cvprblue}{rgb}{0.21,0.49,0.74}
\usepackage[pagebackref,breaklinks,colorlinks,allcolors=cvprblue]{hyperref}

\title{Learning biologically relevant features in a pathology foundation model using sparse autoencoders}

\author{%
Nhat Minh Le$^{1}$ \quad Ciyue Shen$^{1}$ \quad Neel Patel$^1$ \quad Chintan Shah$^1$ \quad Darpan Sanghavi$^1$ \\
Blake Martin$^1$ \quad Alfred Eng$^1$ \quad Daniel Shenker$^1$ \quad Harshith Padigela$^1$ \quad Raymond Biju$^1$ \\
Syed Ashar Javed$^1$ \quad Jennifer Hipp$^1$ \quad John Abel$^1$ \quad Harsha Pokkalla$^1$ \quad Sean Grullon$^1$ \quad Dinkar Juyal$^1$ \\
$^1$PathAI Inc, Boston, USA  \\
\texttt{nhat.le@pathai.com}\
}
\begin{document}
 \maketitle
 \begin{abstract}
Pathology plays an important role in disease diagnosis, treatment decision-making and drug development. Previous works on interpretability for machine learning models on pathology images have revolved around methods such as attention value visualization and deriving human-interpretable features from model heatmaps.
Mechanistic interpretability is an emerging area of model interpretability that focuses on reverse-engineering neural networks. Sparse Autoencoders (SAEs) have emerged as a promising direction in terms of extracting monosemantic features from polysemantic model activations. In this work, we trained a Sparse Autoencoder on the embeddings of a pathology pretrained foundation model.
We found that Sparse Autoencoder features represent interpretable and monosemantic biological concepts. In particular, individual SAE dimensions showed strong correlations with cell type counts such as plasma cells and lymphocytes. These biological representations were unique to the pathology pretrained model and were not found in a self-supervised model pretrained on natural images. We demonstrated that such biologically-grounded monosemantic representations evolved across the model's depth, and the pathology foundation model eventually gained robustness to non-biological factors such as scanner type. The emergence of biologically relevant SAE features was generalizable to an out-of-domain dataset. Our work paved the way for further exploration around interpretable feature dimensions and their utility for medical and clinical applications.
\end{abstract}
    
 \section{Introduction}
\label{sec:intro}

\subsection{Mechanistic Interpretability}

Artificial Intelligence (AI) has made significant strides in various domains, including healthcare and pathology. As these systems become more complex and widely adopted, understanding their internal mechanisms becomes crucial for ensuring reliability, addressing biases, and fostering trust. This paper focuses on the application of mechanistic interpretability (MI) techniques, particularly sparse autoencoders (SAEs), to neural networks used in pathology.

Mechanistic interpretability aims to study neural networks by reverse-engineering them, providing insights into their internal workings \cite{olah2022MechInterpConcepts, cammarata2020circuit, elhage2021mathematical, bereska2024mechanistic}. This approach is particularly relevant in pathology, where understanding the decision-making process of AI systems can have significant implications for patient care and diagnostic accuracy. In the MI paradigm, ``features'' are defined as the fundamental units of neural networks, and ``circuits'' are formed by connecting features via weights \cite{cammarata2020circuit}. This conceptualization allows researchers to dissect complex neural networks and understand how they process and represent information.

According to the Superposition Hypothesis \cite{elhage2022superposition, olah2020Circuits}, a neuron can be polysemantic, i.e., it can store multiple unrelated concepts. Consequently, a neural network can encode more features than its number of neurons. This concept is particularly intriguing in the context of pathology, where complex visual patterns and subtle tissue variations must be recognized and interpreted.

SAEs have been used in NLP \cite{bricken2023monosemanticity, cunningham2023sparseautoencodershighlyinterpretable} to achieve a more monosemantic unit of analysis compared to the model neurons. In vision datasets, SAEs trained on layers of convolutional neural nets have uncovered interpretable features such as curve detectors \cite{gorton2024missingcurvedetectorsinceptionv1, cammarata2020curve}. Various improvements to SAEs have been suggested, including k-sparse \cite{ksparseautoencoders} and gated sparse \cite{gatedsparsesae} autoencoders, and using JumpReLU \cite{jumprelu} instead of ReLU as the activation function. 

In Large Language Models (LLMs), MI has been used to understand phenomena such as in-context learning \cite{olsson2022context}, grokking \cite{nanda2023progress}, and uncovering biases and deceptive behavior \cite{templeton2024scaling}. While these studies primarily focus on language models, their insights may have implications for image-based AI systems used in pathology. The Universality Hypothesis \cite{olah2020Circuits} posits that similar features and circuits are learned across different models and tasks. However, other studies \cite{chughtai2023toy} have found mixed evidence for this claim. Understanding the extent of universality in neural networks could have significant implications for the transferability and generalizability of AI systems in pathology across different types of analyses or tissue samples.

\subsection{Interpretability in Pathology}

Histopathology, often used interchangeably with pathology, is the diagnosis and study of diseases through microscopic examination of cells and tissues. It plays a critical role in disease diagnosis and grading, treatment decision-making, and drug development \cite{walk2009role, MADABHUSHI2016170}. Digitized whole-slide images (WSIs) of pathology samples can be gigapixel-sized, containing millions of areas of interest and biologically relevant entities across a wide range of characteristic length scales. 

Machine learning (ML) has been applied to pathology images for tasks such as segmentation of biological entities, classification of these entities, and end-to-end weakly supervised prediction at a WSI level \cite{bulten2020automated, campanella2019clinical, wang2016deep}. Work on interpretability in pathology has focused on assigning spatial credit to WSI-level predictions \cite{javed2022additive, lu2020data}, computing human-interpretable features from model output heatmaps \cite{diao2021human}, and visualization of multi-head self-attention values on image patches \cite{chen2024uni}.

Foundation Models (FMs) are promising for pathology as they can take advantage of large amounts of unlabeled data to build rich representations which can be easily adapted for downstream tasks in a data-efficient manner \cite{lunit, rudolfv, virchow, phikon, chen2024uni}. The diversity of pre-training data powers these models to generate robust representations, enabling them to generalize better than individual task-specific models trained on smaller datasets.  Additionally, these models can be used as a universal backbone across different tasks, reducing the development and maintenance overhead associated with bespoke task-specific models. 

We believe that histopathology data is a promising area for Mechanistic Interpretability (MI)-based analysis, for the following reasons:

\begin{itemize}
    \item \textbf{Rich and Complex Data:} Unlike object-centric image datasets, a single pathology image patch can contain up to $10^6$ regions of interest (e.g., cell nuclei). The number of active concepts is bounded by underlying biological structures, and identifying every concept can be critical for downstream applications.
    
    \item \textbf{Addressing Batch Effects:} Pathology images are susceptible to ``batch effects'', where models may learn spurious features instead of relevant morphology-related features. This issue arises from high-frequency artifacts and systematic confounders in image acquisition \cite{Howard2020}. MI can help disentangle biological content from incidental attributes, leading to more robust models for real-world applications.
    
    \item \textbf{Enabling Precise Interventions:} A bottom-up understanding of feature contributions to predictions can enable modeling of useful interventions at increasing levels of complexity. This ranges from activation-based methods \cite{vig2020causal, chan2022causal} to text-based interventions, such as predicting tissue changes in response to drug administration.
    
    \item \textbf{Enhancing Model Transparency:} MI can provide insights into the decision-making process of AI systems in pathology, potentially improving their interpretability and trustworthiness in clinical settings.
    
    \item \textbf{Facilitating Novel Discoveries:} By uncovering the internal mechanisms of AI models trained on pathology data, MI may lead to new biological insights or hypotheses that were not apparent through traditional analysis methods.
\end{itemize}

\subsection{Summary of Contributions}

This work presents an interpretability analysis of the embedding dimensions derived from a vision foundation model trained on histopathology images. We employ SAEs on the model embeddings to uncover monosemantic representations of biologically-relevant and human interpretable features. Our study provides the first detailed characterization of the image attributes represented within embedding dimensions of a pathology foundation model and how they evolve through the model’s depth.

The main contributions of our work are as follows:
\begin{itemize}
    \item We demonstrate that individual dimensions in the embedding space encapsulate complex, higher-order concepts through polysemantic combinations of fundamental characteristics like cell appearance and nuclear morphology.
    \item We train a sparse auto-encoder (SAE) to disentangle polysemantic embedding dimensions, revealing a sparse dictionary of interpretable features that represent cell and tissue characteristics, geometric structures, and image artifacts.
    \item We investigate the features identified by the SAEs and show that specific SAE dimensions are correlated with counts of key cell types in pathology (plasma cells, cancer cells, lymphocytes, fibroblasts, macrophages). 
    \item These cell-type specific dimensions are found in the last layers, while SAE dimensions from the earlier layers mainly correlate with color features, providing evidence that the model gains invariance to spurious factors for downstream pathology tasks.
    \item We demonstrate that the SAE representations are robust and can be generalized across metadata such as disease, stain and scanner type.
\end{itemize}

\begin{figure*}[h]
\vskip 0.2in
\begin{center}
\centerline{\includegraphics[width=1.5\columnwidth]{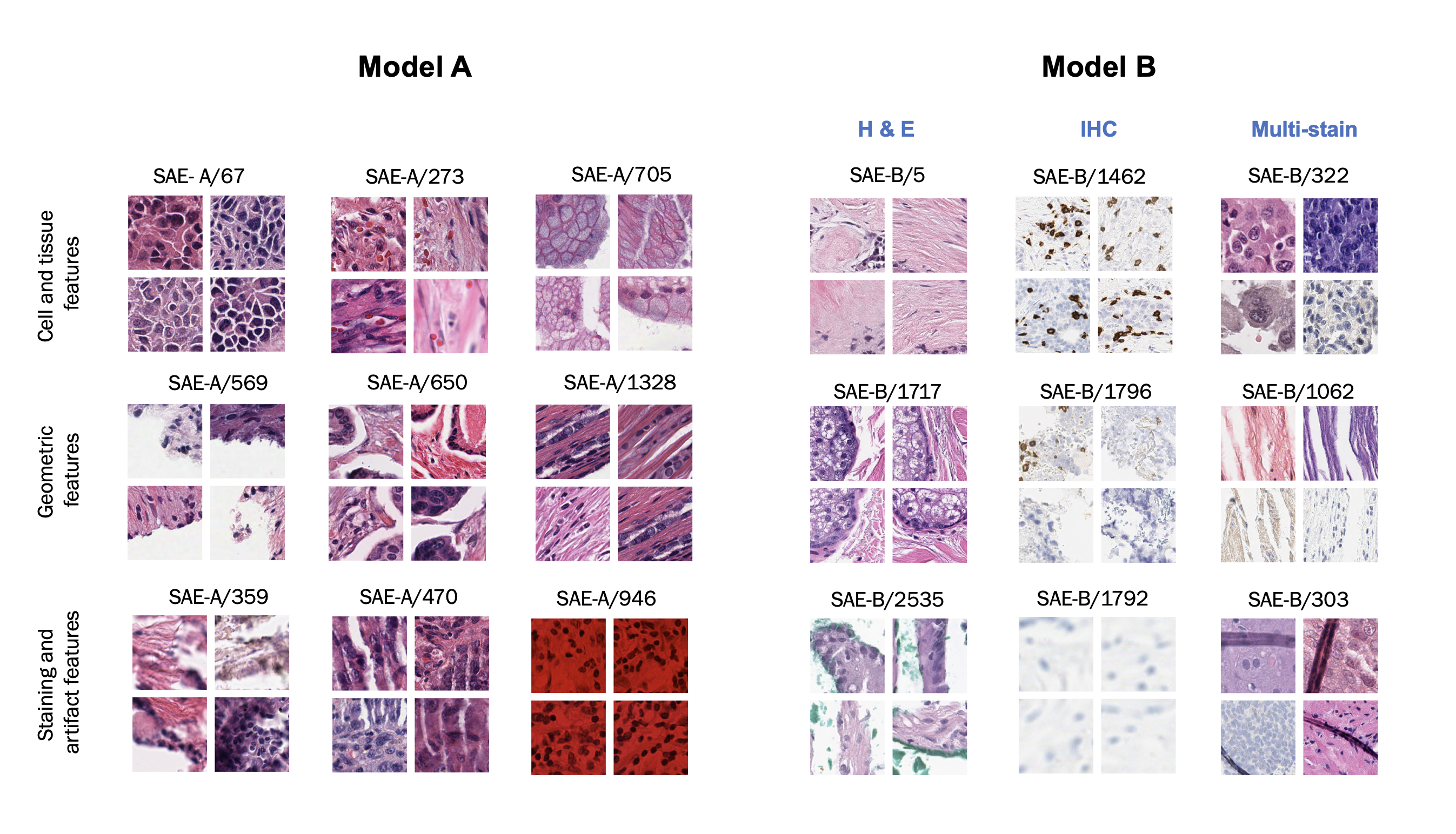}}
\caption{Feature visualization of SAE hidden dimensions revealed interpretable dictionary of pathology features. For each SAE hidden dimension of model A (trained on the TCGA dataset) and model B (trained on the 1M dataset), 4 out of the top 16 images that activated that dimension were visualized. Manual examination revealed interpretable features represented by these dimensions. For model A, these include cell and tissue features specific to H \& E stain (top row: poorly differentiated carcinoma with distinct cell separation, red blood cells, mucin); geometric features (middle row: edge of tissue, clefting between cancer and stroma, diagonal fibers); staining and artifact features (bottom row: blur, sectioning artifact, red stain). For model B, some SAE dimensions were specific to H \& E stain (first column: collagen-enriched fibroblasts, circular clusters of tumor cells, surgical ink), some were specific to IHC stain (second column: stained lymphocytes, edge of tissue, blur), and others generalized across stains (third column: large cancer cells, vertical structures, tissue folds).}
\label{fig:sae-features}
\end{center}
\vskip -0.2in
\end{figure*}

\section{Method} \label{sec:Method}

\subsection{Datasets}
We used three datasets for experimentations. For the training dataset, we used 1.1 million image patches (1M dataset) (224 x 224 pixels at a high resolution, 0.25 microns per pixel) including both haematoxylin \& eosin (H \& E) and immunohistochemistry (IHC) stains, sampled from the train set of `PLUTO' - a pathology pretrained foundation model \cite{juyal2024pluto}, covering oncology, IBD (inflammatory bowel disease) and MASH (metabolic dysfunction-associated steatohepatitis).

Two different datasets were mainly used for evaluation. The first (TCGA dataset) comprised three publicly available TCGA (The Cancer Genome Atlas) \cite{weinstein2013cancer} datasets containing H $\&$ E-stained histology images from three organs: breast (TCGA-BRCA), lung (TCGA-LUAD), and prostate (TCGA-PRAD). We selected 951, 493 and 488 WSIs from these datasets respectively for the analysis. This dataset was used in the early experimentation of SAE training. The second dataset (CPTAC) comprised two publicly available CPTAC (Clinical Proteomic Tumor Analysis Consortium) 
datasets containing H $\&$ E-stained histology images from two cancer types: cutaneous melanoma (CPTAC-CM, 256 WSIs), and head and neck cancer (CPTAC-HNSCC, 228 WSIs). The two evaluation datasets were chosen from different data sources and purposely from different organ types to maximize the data diversity.

\subsection{Embedding extraction}
All the images for the train and evaluation datasets were passed through a frozen ViT-S encoder taken from `PLUTO'. For each image patch, we extracted 384-dimensional embedding vectors corresponding to the CLS token residual stream in layers 1-12 with 12 as the output layer.
For baseline comparison, we used a self-supervised vision transformer DINO \cite{DBLP:journals/corr/abs-2104-14294}.

\subsection{Biological and color feature extraction}
\label{susbsec:feature_extraction}

To investigate the representation of cellular features in SAE dimensions, we deployed PathExplore, a machine-learning model (PathExplore is for research use only. Not for use in diagnostic procedures) \cite{markey2023abstract, Abel2024}, on the evaluation datasets. On each slide, the model detected and classified cell types including cancer cells, lymphocytes, macrophages, fibroblasts, plasma cells. The count and density of each cell class on each image were computed.
In addition, color features, including gray-scale intensity, LAB colorspace, and saturation, were extracted in each image patch by taking the average or standard deviation of feature values across all the pixels in the image at its original resolution.

\subsection{Sparse autoencoder training} \label{subsec:sae_training}
The SAEs were trained using a loss function given by $\frac{1}{k} (\sum_{i=1}^{k}||\textbf{x}_i - \hat{\textbf{x}_i}||_2 + \lambda\sum_{i=1}^{k}||\textbf{f}_i||_1)$, where $k$ is the batch size, $\textbf{x}_i$ and $\hat{\textbf{x}}_i$ are the raw and reconstructed embeddings, and $\textbf{f}_i$ are the learned features of image $i$ \cite{bricken2023monosemanticity, sparse_autoencoder}. Dead neuron resampling was implemented to reduce the fraction of dead neurons \cite{bricken2023monosemanticity, sparse_autoencoder}. We used Adam optimizer with a learning rate of 0.001, expansion factors of 1, 8, 16, 32; and L1-penalty weight in {0.001, 0.004, 0.006, 0.008, 0.01}. Results were reported for models with an expansion factor of 8 and L1-penalty weight of 0.004.

\section{Training a sparse autoencoder on PLUTO embeddings reveals interpretable features}

\subsection{Interpretability analysis of PLUTO embeddings}
We first manually inspected each of the 384 dimensions of the PLUTO embedding space to determine if they represent singular features of the images. For each dimension, we randomly sampled 5 patches that have the lowest 5\% and the highest 5\% activation values across the TCGA-BRCA dataset (see Supplementary section).

The embedding dimensions tended to encode multiple image characteristics. For example, dimension 27 was more activated for larger cells (compared to smaller cells), purple background (compared to red background), and non-elongated cell shapes. Dimension 118 was more activated for mucinuous and round structure and less activated for fibrous structures.

By visual inspection, most embedding dimensions encoded a combination of these cellular, tissue and background-stain related characteristics, suggesting a polysemantic representation of atomic properties. Certain combinations of the atomic properties corresponded to complex concepts that were relevant to pathology, such as the distinction between cancer epithelium and stroma tissue (captured in dimension 27 and 147), or the presence of red blood cells (captured in dimension 239). However, the polysemantic features represented in these dimensions prevented interpretability analysis of these dimensions.

Sparse autoencoder models were fit to the CLS token embedding from the output layer (layer 12) of the training dataset (1M dataset). Training the SAE on this dataset (including multiple organs, stains and cell types) leads to generalizable representation of useful features in the embedding dimensions of the model. We visualized the images that have the highest activation value for a given SAE dimension. This revealed highly interpretable features, as shown in Figure \ref{fig:sae-features}, including cell and tissue features such as poorly differentiated carcinoma, geometric structures such as vertical fibers, and staining and artifact features.

With the use of the diverse training set, individual dimensions of the trained SAE model exhibited robust representations, where single dimensions represented the same features regardless of stain type. Consistent with this, 247/3072 dimensions (8.0\%) had representations of both H \& E and IHC stains in the top 100 activating patches, and some of these dimensions represented interpretable concepts across stain types (Figure \ref{fig:sae-features}, rightmost column). 374/3072 dimensions (12.2\%) were H \& E-specific while 1451/3072 dimensions (47.2\%) were IHC-specific. This result showed that when trained with diverse datasets, SAE dimensions can represent stain-specific features and exhibit cross-stain generalization.

\begin{figure*}[h]
\vskip 0.2in
\begin{center}
\centerline{\includegraphics[width=1.5\columnwidth]{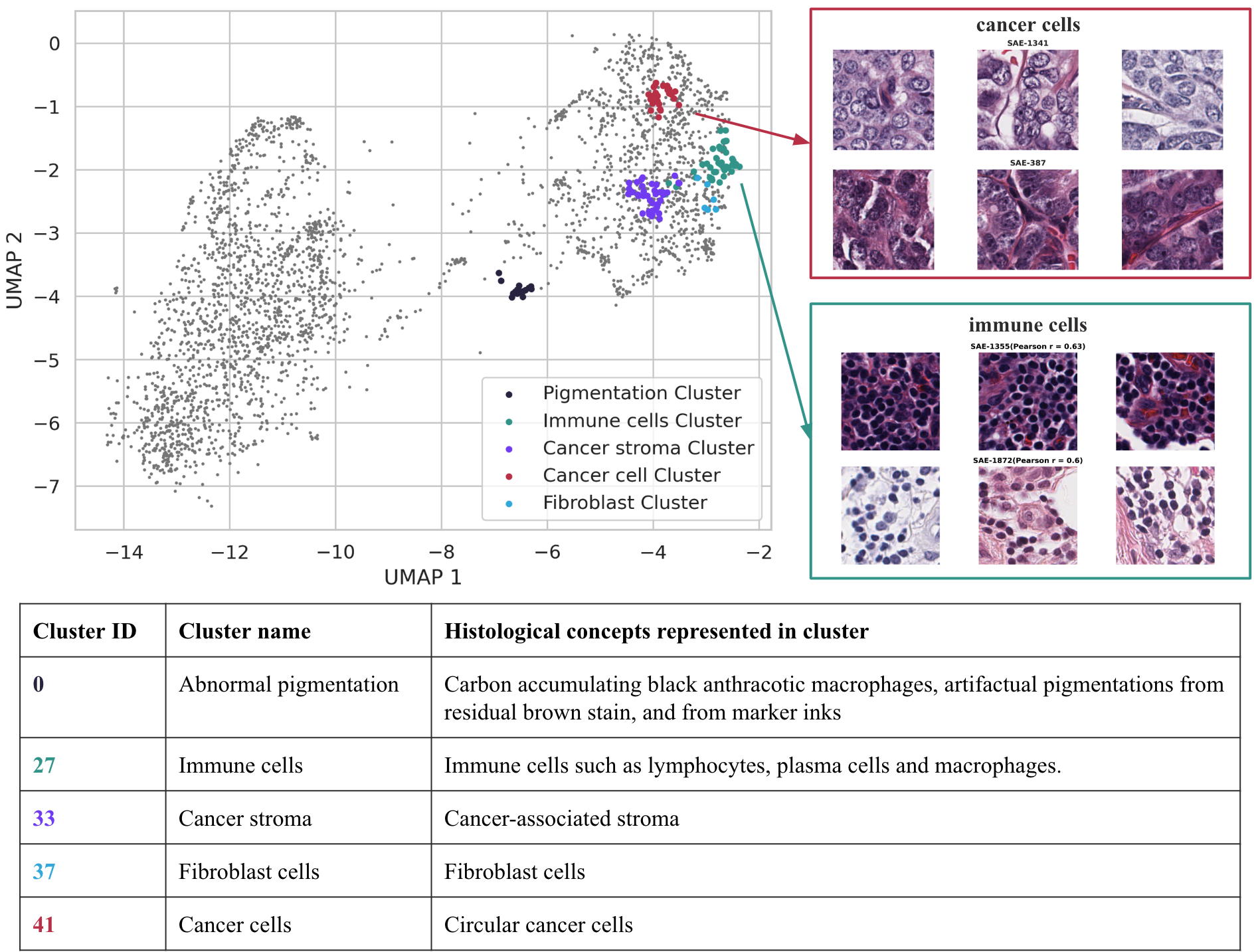}}
\caption{UMAP of 3072 SAE dimensions from model trained on the 1M dataset. Feature clusters were identified by HDBSCAN and were interpreted by manual inspection. Several clusters clearly associated with histological concepts were highlighted. For cancer and immune cell clusters, visualizations of top 3 patches that maximally activate the SAE dimension were shown.}
\label{fig:umap_viz}
\end{center}
\vskip -0.2in
\end{figure*}

Training on the diverse dataset reduced the fraction of dead neurons in the SAE intermediate layer compared to a model that was trained only on the TCGA dataset. Similar to previous work for natural language \cite{bricken2023monosemanticity}, we identified a cluster of ``ultra-sparse" features that activated for very few images ($<0.1\%$ of the dataset). The fraction of these ultra-sparse features were reduced with the incorporation of more diverse training data for the model trained on the 1M dataset ($20\%$) compared to the model trained on TCGA dataset ($88\%$) (see Supplementary section). For subsequent analyses, we used the SAEs trained on the 1M dataset, and used the TCGA dataset as a held-out evaluation dataset.

\subsection{PLUTO SAE dimensions represent interpretable pathology-relevant concepts}
Using the TCGA evaluation dataset, we performed unsupervised clustering on the UMAP representations of the SAE dimensions using HDBSCAN, following the analysis strategy of \cite{bricken2023monosemanticity} (Figure \ref{fig:umap_viz}). To understand the meanings of some of the clusters, we manually examined image patches activating the SAE dimensions within each cluster.

Of the 139 clusters obtained using HDBSCAN, we found clusters (Figure \ref{fig:umap_viz}) containing SAE features correlated with unique histological concepts such as immune cell presence (Cluster 27), cancer stroma (Cluster 33), fibroblast cells (Cluster 37) and circular cancer cells (Cluster 41). Notably, cluster 0 features were associated with abnormal pigmentation, such as carbon accumulating black anthracotic macrophages (SAE-1745) as well as artifactual pigmentations from residual brown stain (SAE-2034) and from marker ink (SAE-2842) (see Supplementary section).

\subsection{Comparison to non-pathology ViT model}
To investigate whether the pathology-relevant SAE features emerged due to the pre-training method of PLUTO on pathology images, we evaluated our SAE methodology on a baseline ViT-S that matched PLUTO's design. This ViT-S was trained on ImageNet-1k using the self-supervised DINO method, and was obtained from the timm library \cite{caron2021emerging,dosovitskiy2020vit,rw2019timm}. We chose the same input patch size as PLUTO and SAE training methodology as in section \ref{sec:Method} to ensure fair comparison.

\begin{figure}[H]
    \centering
    \includegraphics[width=\linewidth]{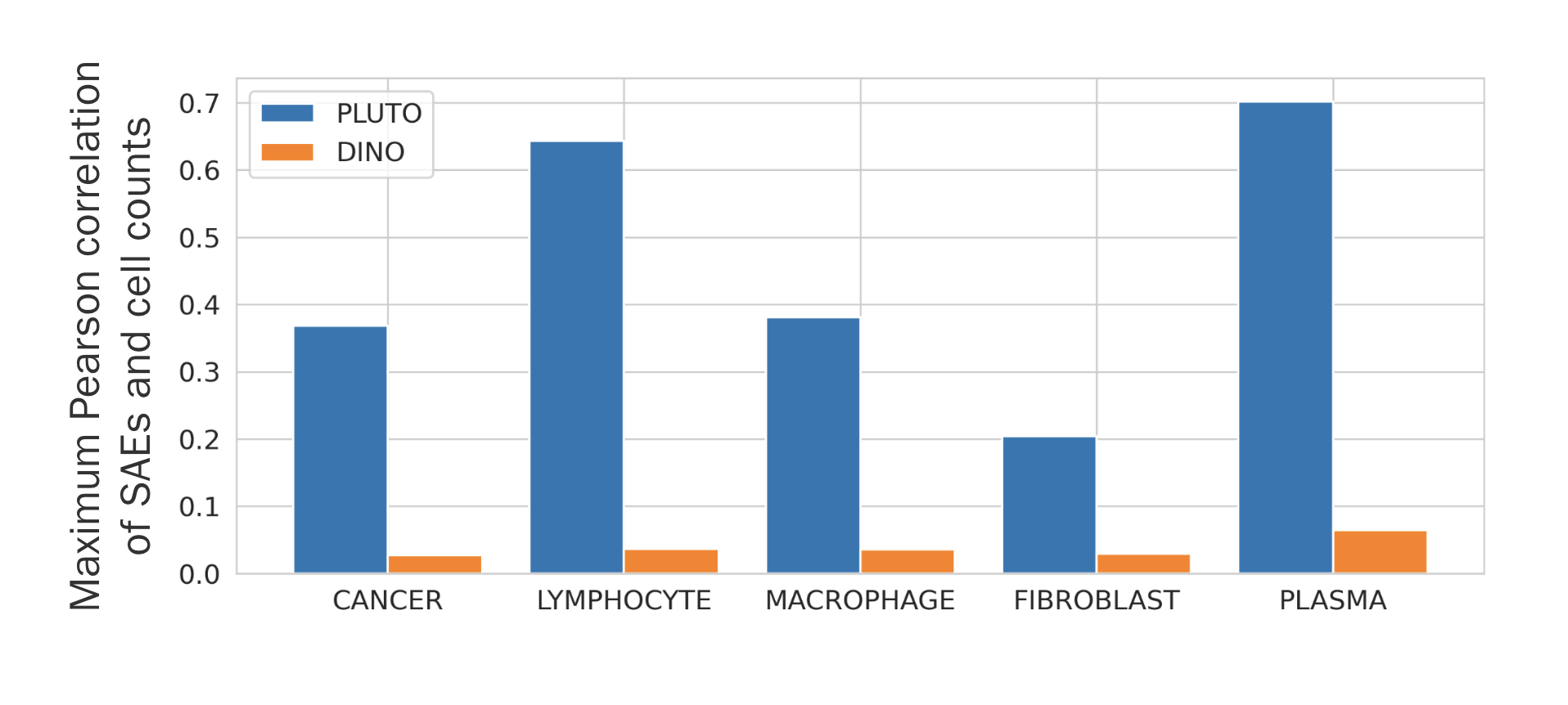}
    \caption{Pearson correlations of SAE dimensions of PLUTO and DINO models with counts of pathology-relevant cell types, showing much higher correlations of the PLUTO SAE dimensions with the cell count features.}
    \label{fig:baseline}
\end{figure}

To compare the SAE representations of PLUTO and DINO, we used human-interpretable features (HIFs) \cite{diao2021human} quantifying tumor microenvironment characteristics such as counts of cancer cells, plasma cells, lymphocytes, macrophages and fibroblasts (see Section \ref{susbsec:feature_extraction}). We first calculated the Pearson's correlation ($\rho$) of the SAE activation values of the PLUTO with five human-interpretable features representing cell counts. We identified SAE dimensions from PLUTO with the highest correlation with each cell count HIF: plasma cells ($\rho$ = 0.7), lymphocytes ($\rho$) = 0.63), cancer cells ($\rho$) = 0.37), macrophages ($\rho$) = 0.38), and fibroblasts ($\rho$) = 0.21).

In contrast to the SAE dimensions of PLUTO, SAE dimensions of DINO showed weak association with pathology-relevant concepts. SAE dimensions of this model showed poor correlations with the counts of different cell types (Figure \ref{fig:baseline}). These differences in the representations of the two models demonstrated the value of training on pathology-specific data.

\subsection{Evaluation of SAE monosemanticity using pathology-relevant cellular features}
\label{subsec:sae_monosemanticity}

\begin{figure*}[h]
    \centering
    \includegraphics[width=0.70\linewidth]{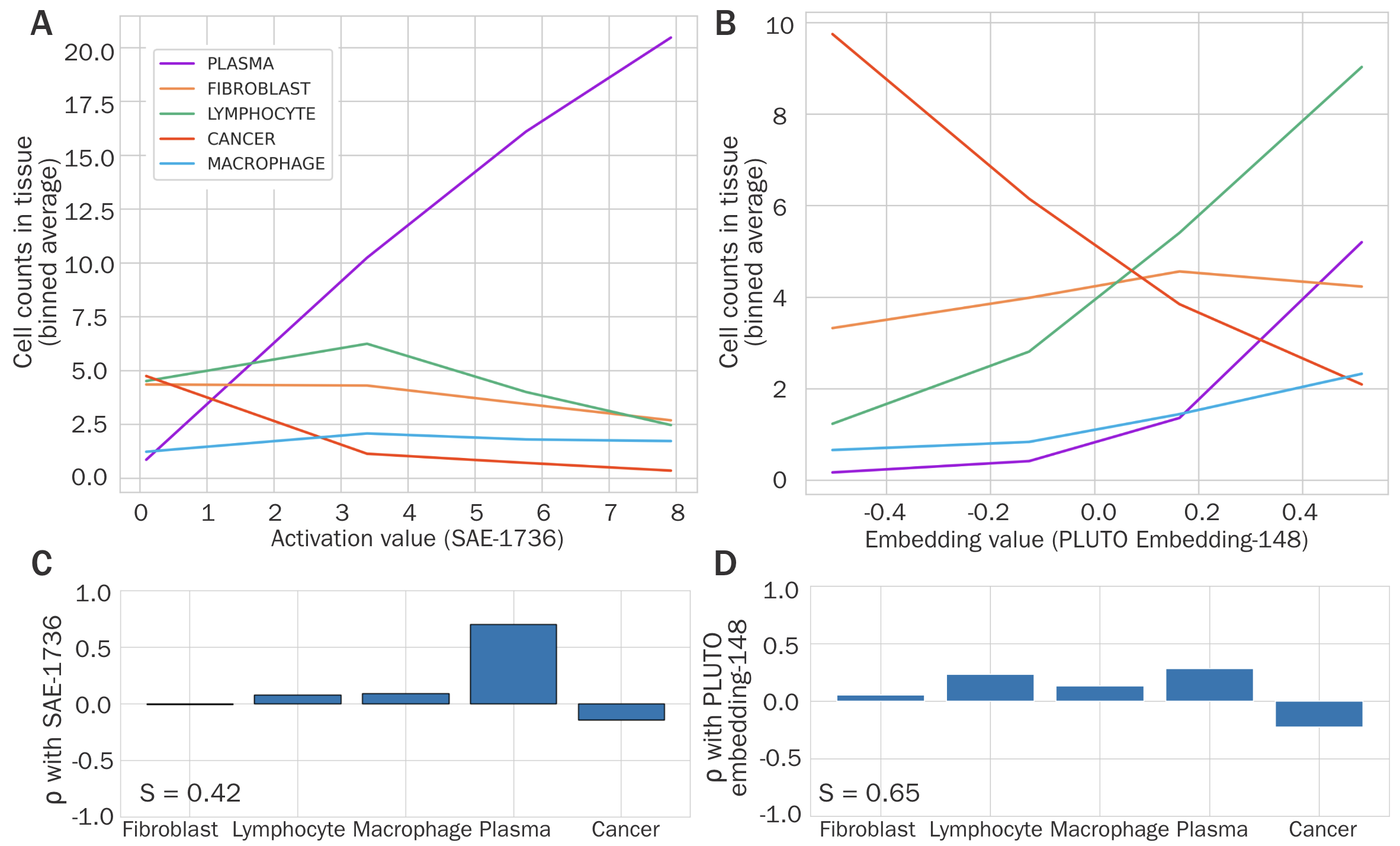}
    \caption{SAE-1736 monosemantically encoded plasma cell-specific information. Top panels show the average cell counts across bins of (A) SAE-1736 activation values, and (B) PLUTO dimension 148. Average plasma cell counts (shown in purple) increased linearly with increasing SAE-1736 activation values, while counts of other cell types decreased or remained constant. In contrast, counts of lymphocytes, macrophages, and plasma cells all increased monotonically with increasing PLUTO-148 feature values. C) Correlation between SAE-1736 activation and counts of five cell types, showing monosemantic correlation with only the plasma cell counts. D) Same as C, but for PLUTO dimension 148}
    \label{fig:sae-lineplot}
\end{figure*}

To evaluate the monosemanticity of the SAE dimensions, we investigated whether the SAE with highest correlation with the counts of specific cell types also associated with other cell types. SAE-1736, which exhibited a strong correlation with plasma cell counts, showed minimal correlation ($\rho < 0.1$) with other cell types. Images with the highest activation values for SAE-1736 consistently demonstrated a high presence of plasma cells and captured specific histological features, such as eccentric nuclei surrounded by pale blue cytoplasm. The linear relationship between SAE-1736 activation and plasma cell counts was further illustrated in Figure~\ref{fig:sae-lineplot}A,C. As the average SAE-1736 activation increased, plasma cell counts rose linearly, while the counts of other cell types remained constant or decreased.

In contrast, no such monosemantic feature was found in the PLUTO embedding space. The strongest plasma cell-associated PLUTO dimension, 148, exhibited only a moderate correlation with plasma cell counts ($\rho$ = 0.29) and was also correlated with the presence of other cell types, as shown in Figure~\ref{fig:sae-lineplot}B and D. To quantify monosemanticity with respect to these five cellular concepts, we defined a probability distribution $p_i = \frac{|\rho_i|}{\sum_j |\rho_j|}$ where $\rho_1, ...\rho_5$ are the Pearson correlations with the individual cell class densities. Monosemantic features would result in a low entropy $S$ of the distribution. We found that SAE-1736 has a low entropy ($S=0.42$), while PLUTO dimension 148 has higher entropy ($S=0.65$, close to the maximum possible entropy $S_{max}=0.70$). This suggested that SAE dimensions both show stronger correlations with biological concepts (plasma cells), and a higher degree of monosemanticity than original PLUTO embedding dimensions.

\section{Emergence of monosemanticity across PLUTO layers}
\label{sec:sae_representations}

Previous work on convolutional networks \cite{zeiler2013visualizingunderstandingconvolutionalnetworks} and vision transformers \cite{ghiasi2022visiontransformerslearnvisual} suggested emergence of complex features in downstream layers of deep networks. Via SAEs, we investigated how the representation of pathology-relevant concepts evolve across layers of PLUTO.

\begin{figure*}
    \centering
    \includegraphics[width=0.9\textwidth]{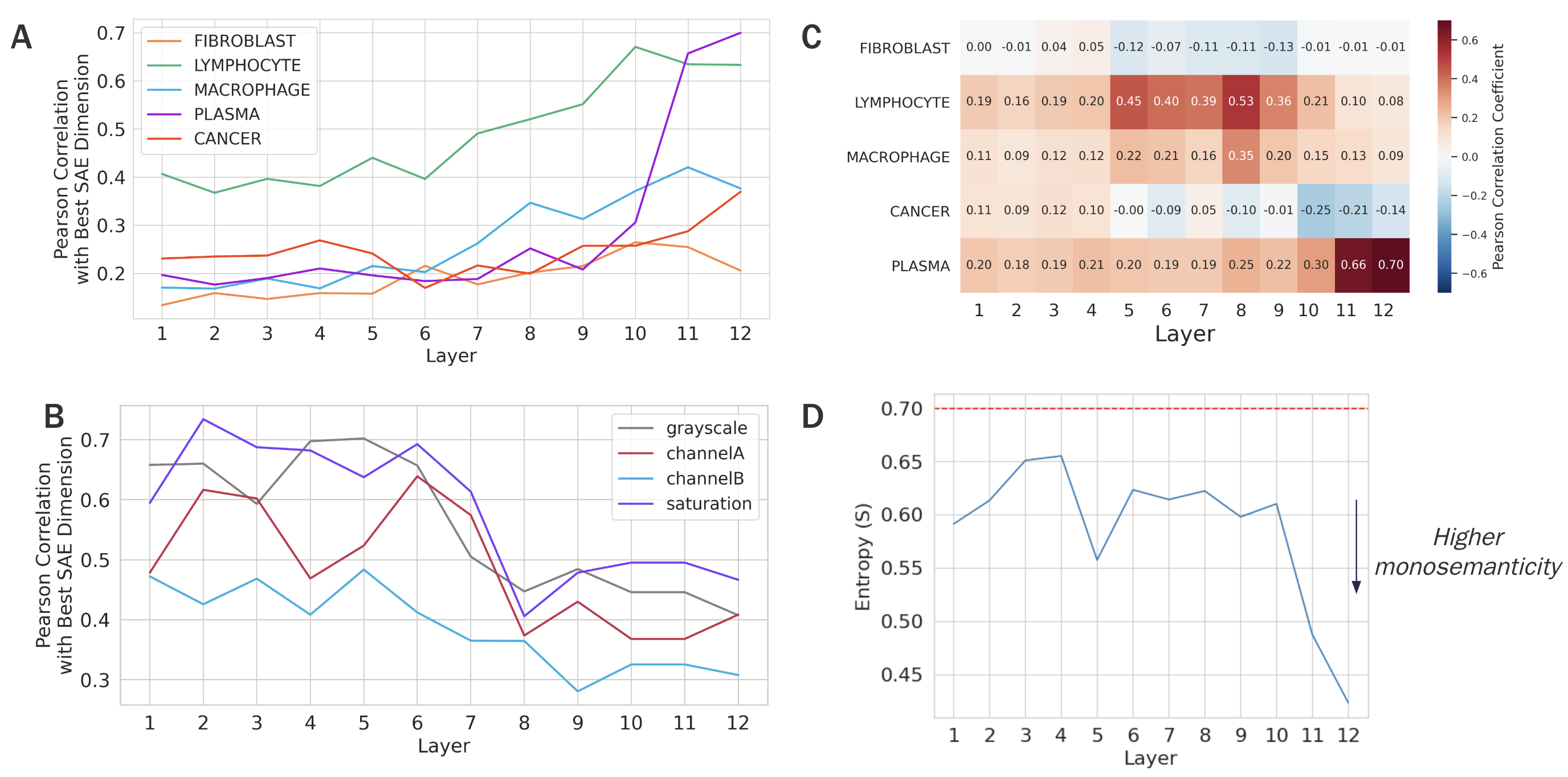}
    \caption{Monosemanticity emerged in later layers of PLUTO A) Correlation of cell count features with dimensions of SAE models trained on the embeddings of PLUTO across layers. In each layer, the five SAE dimensions with the highest correlation with the counts of each of the five cell classes were plotted. B) Correlation of color features with dimensions of SAE models trained on the embeddings of different layers of PLUTO C) For each layer, we found the SAE dimension with the highest correlation with count of plasma cells. We then measured monosemanticity of that dimension by calculating the correlation with other cell type counts. D) Entropy of the best plasma-cell SAE for each layer with respect to the five cell types (lower entropy implies higher monosemanticity). Red dotted line represents maximum possible entropy ($S_{max} = 0.70$) }
    \label{fig:sae_monosemanticity}
\end{figure*}

For this investigation, we extracted the embeddings of the CLS token in every layer of PLUTO and trained separate SAEs on these embeddings. $\rho$ for maximally correlated SAEs per cell type for each layer was shown in Figure\ref{fig:sae_monosemanticity}A. At earlier layers (L1 to L6), SAE dimensions correlated with low-level color features such as intensity, hue and saturation (Figure \ref{fig:sae_monosemanticity}B). Correlations of these SAE dimensions with cellular features were low ($\rho < 0.5$ for lymphocytes and $\rho<0.3$ for the other four cell types).

At later layers (L7 to L12), association of SAE dimensions with color features decreased, while association with cell features increased, suggesting that the feature representations in these layers were biologically meaningful while being invariant to lower level features. More specifically, the SAE representation of plasma cells did not emerge until layer 10 (Figure \ref{fig:sae_monosemanticity}A). 

Investigating how plasma cell monosemanticity emerged across layers, we observed that the maximally correlated plasma cell SAEs for earlier layers had higher correlation with lymphocyte counts (e.g. Layer 5 SAE-32 $\rho$ = 0.45, Layer 9 SAE-100 $\rho$ = 0.36, Figure \ref{fig:sae_monosemanticity}C). In other words, embeddings from earlier PLUTO layers could not produce monosemantic SAEs related to plasma cell presence, but might be capturing some simpler and generic characteristics shared between lymphocytes and plasma cells, such as darkly stained nuclei. This was consistent with our observations on correlation of these SAE dimensions with color features. We found a decrease in the entropy of the SAE dimension that showed the strongest association with plasma cell counts (Figure \ref{fig:sae_monosemanticity}D), suggesting increased monosemanticity in later layers.

\begin{figure}
    \centering
    \includegraphics[width=0.4\textwidth]{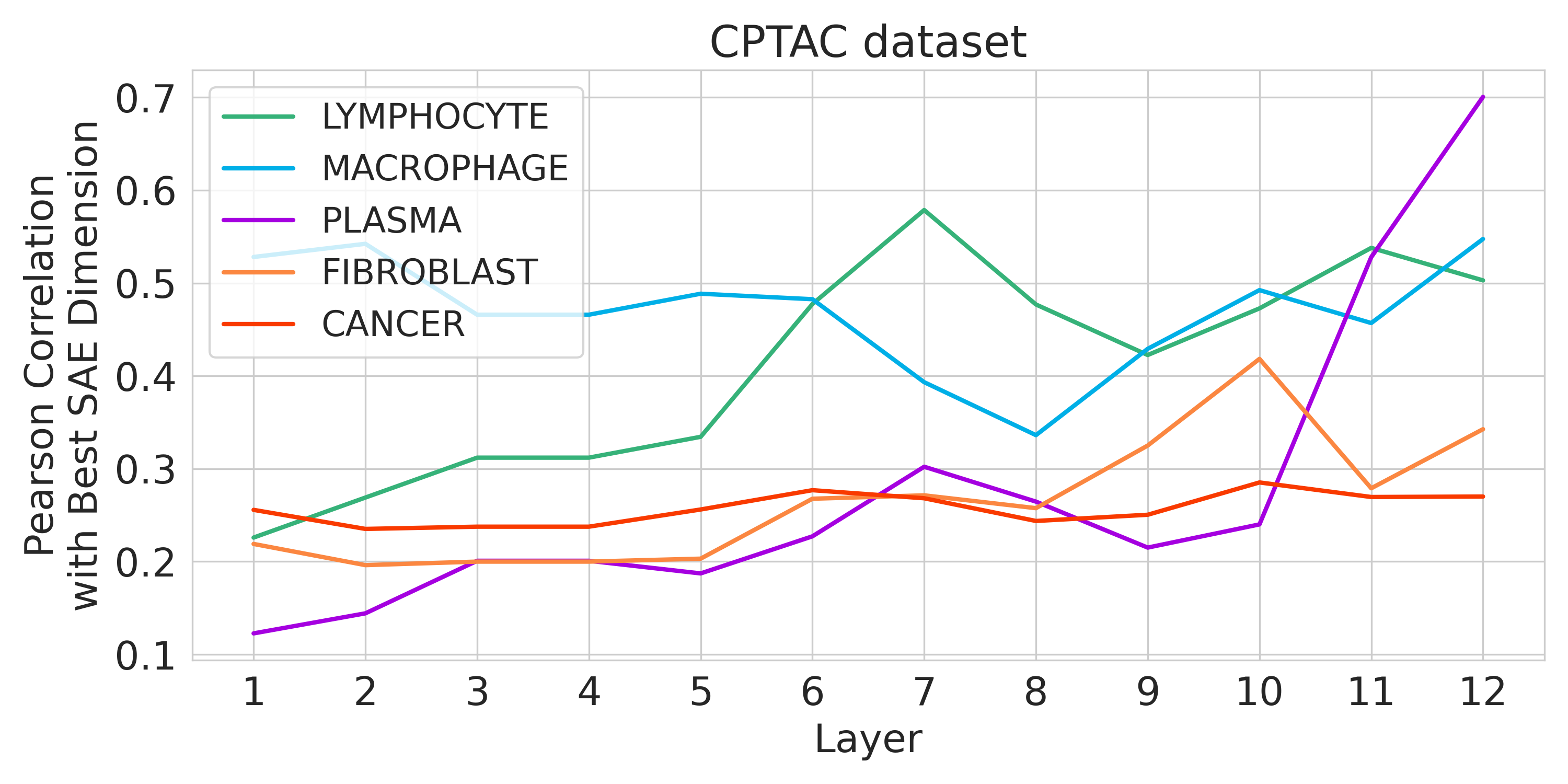}
    \caption{Maximum correlations of SAE dimensions with cellular features in CPTAC dataset. Method for computing the Pearson correlation at each layer is the same as Figure \ref{fig:sae_monosemanticity}A}
    \label{fig: cptac}
\end{figure}

\section{Robustness of SAE representations across domains}
\subsection{SAE correlations on an out-of-domain evaluation dataset}

Finally, we verified that our previous results extended across other domains of pathology images. We extracted embeddings and deployed SAEs on the CPTAC dataset, which covers two different oncology indications. We performed similar correlation analysis between SAE dimensions and cell features. We confirmed the following discoveries (Figure \ref{fig: cptac}): 1. representations of biologically relevant cellular features by SAE evolved across PLUTO layers, and 2. the monosemantic representation of plasma cell by SAE-1736 was robust across different domains of pathology images.

\subsection{Robustness of PLUTO to scanners and stains}
The low correlation of the SAE dimensions with color features suggested that PLUTO learned a representation of the pathology-relevant concepts that was robust to sources of variation such as stain types and scanner type which can have a large impact on model performance of downstream tasks.

\begin{figure}
    \centering
    \includegraphics[width=0.43\textwidth]{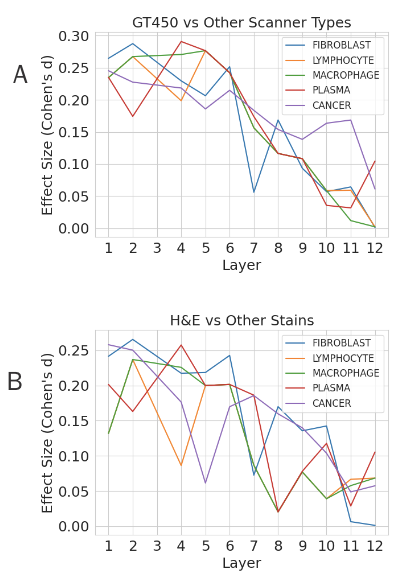}
    \caption{Robustness of cell-type specific SAE dimensions to scanners and stains. In each layer, we identified five SAE dimensions that showed the highest Pearson correlation with counts of fibroblasts, lymphocytes, macrophages, plasma cells and cancer cells. The difference in activation of those SAE dimensions were computed between patches from (A) GT450 versus other scanner types, or (B) H\&E versus other stains. Effect sizes were quantified by Cohen's d metric.}
    \label{fig:scanner_stain_robustness}
\end{figure}

To investigate whether the cell-type specific SAE dimensions as identified in sections \ref{subsec:sae_monosemanticity} and \ref{sec:sae_representations} were robust to scanners, we computed differences in these SAE activations between the images scanned with GT450 scanners and all other images (quantified by Cohen's d, Figure \ref{fig:scanner_stain_robustness}A). The Cohen's d metric was moderate in the earlier layers and decreased to close to zero at the last layer, showing that the cell-type specific dimensions identified above did not show systematic scanner-specific variations. A similar analysis was done on H\&E and other stains, showing reduced stain-specific variation in the SAE dimensions trained on the last layers of PLUTO (Figure \ref{fig:scanner_stain_robustness}B).

\subsection{Feature universality of SAE dimensions}
Previous work suggested that SAE dimensions are more likely to be useful (representing true monosemantic features in the real world) if they display \textit{universality} (the same feature is discovered across independently trained SAE models \cite{bricken2023monosemanticity}). We examined feature universality of the SAE dimensions of PLUTO by comparing the SAE activations from two models, one trained on the 1M dataset, and one trained on the TCGA dataset. We found that these two models were able to uncover SAE dimensions that captured the same histological concepts. For example, SAE-1736 from the model trained on the 1M dataset and SAE-2541 from the model trained on the TCGA dataset were highly correlated ($\rho$ = 0.96) and both represent abundance of plasma cells; SAE-1745 from model B and SAE-1667 from model A both represent abundance of anthracotic macrophages ($\rho$ = 0.91) (see Supplementary section). These findings demonstrate the universality of the learned SAE features and suggests generalizability of the SAEs.

\section{Limitations and future work}
In this work, we restricted our analysis to a vanilla SAE. We left the application of newer variations such as gated SAE and k-sparse SAE in pathology to future work. Similarly, analysis around how these findings translate to SAEs trained on other pathology foundation models can be the subject of further studies. Another area of exploration can be around performing intervention in the SAE latent space and characterizing its impact on robustness to spurious features.

\section{Conclusion}

We performed an investigation of the features represented in the embedding space of a pathology foundation model. Single embedding dimensions were found to demonstrate polysemanticity in terms of representing higher-order pathology-related concepts composed of atomic characteristics of cellular and tissue properties. Training a SAE enabled the extraction of relatively monosemantic and interpretable features corresponding to distinct biological characteristics, geometric features and image acquisition artifacts. Analysis with human-interpretable features revealed correlations of SAE activations with counts of different cell types. Clustering of SAE dimensions revealed distinct groups corresponding to related and interpretable concepts such as anomalous pigmentation, malignant regions and inflammation. These features demonstrated generalization across multiple stains. 

An in-depth investigation of feature representations of individual layers from the pathology foundation model provided insight on the evolution of these feature across model depth. Generic features of an image, such as colors, were learned early on by the model, while biologically relevant features, such as abundance of individual cell types, emerged at later layers of the model. Consequently, we demonstrated that the model gains invariance to biologically irrelevant features, such as scanner types. This study provided concrete evidence that embeddings extracted from this pathology foundation model were biologically-grounded, facilitating downstream models that are built upon these embeddings for solving pathology-relevant tasks. Overall, investigation of sparse features is a promising direction and motivates further work in discovering explainable, generalizable features of pathology foundation models.

{
    \small
    \bibliographystyle{ieeenat_fullname}
    \bibliography{main}
}
\clearpage
\setcounter{page}{1}
\maketitlesupplementary

\makeatletter
\renewcommand \thesection{S\@arabic\c@section}
\renewcommand\thetable{S\@arabic\c@table}
\renewcommand \thefigure{S\@arabic\c@figure}
\makeatother

\section{Quantification of dead and ultra-low density neurons across SAE models}

For each SAE dimension, we computed the number of dead neurons (neurons that showed no activation across the entire dataset), and the number of ultra-low density features (features where the fraction of active neurons is below 0.1\%).

\begin{table}[h!]
    \vskip 0.15in
    \centering
    \small
    \begin{tabular}{|c|p{1.5cm}|c|c|p{1.5cm}|}
        \hline
        \textbf{Dataset} & \textbf{Expansion factor} & \textbf{Common} & \textbf{Dead} & \textbf{Ultra-low density} \\
        \hline
        TCGA &8  & 9.1\% & 2.7\% & 88.2\%\\
        \hline
        1M&1  & 99.7\% & 0\% & 0.3\% \\
        \hline
        1M&8 & 80\%&0.2\% & 20\%\\
        \hline
        1M&16 &62\% & 0.7\%&37\% \\
        \hline
        1M&32 &46\% & 1.4\%&53\% \\
        \hline
    \end{tabular}
    \caption{Fraction of features belonging to the common, dead or ultra-low density feature groups across models.}
    \label{tab:fraction_dead_neurons}
\end{table}

\vfill\null
\vspace{50cm}
\section{Visualization of individual SAE features within key UMAP clusters}

\begin{figure}[h!]
    \centering
    \includegraphics[width=1\linewidth]{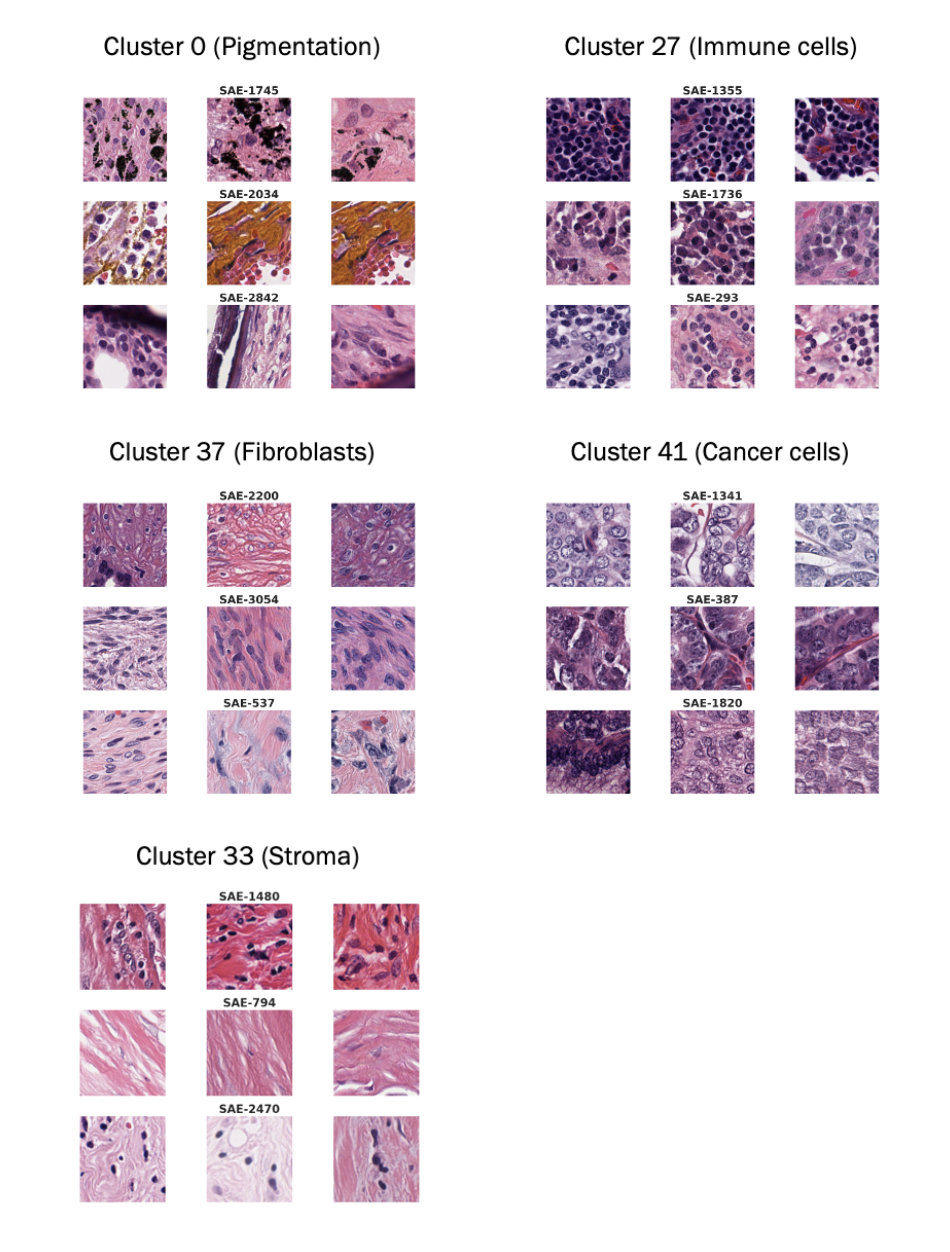}
    \caption{Visualization of features within key clusters identified by the UMAP analysis. For each cluster, each row represents a SAE dimension from that cluster, and shows 3 patches that maximally activate that dimension.}
    \label{fig:clustering_viz}
\end{figure}

\begin{figure*}[h!]
    \centering
    \includegraphics[width=0.7\linewidth]{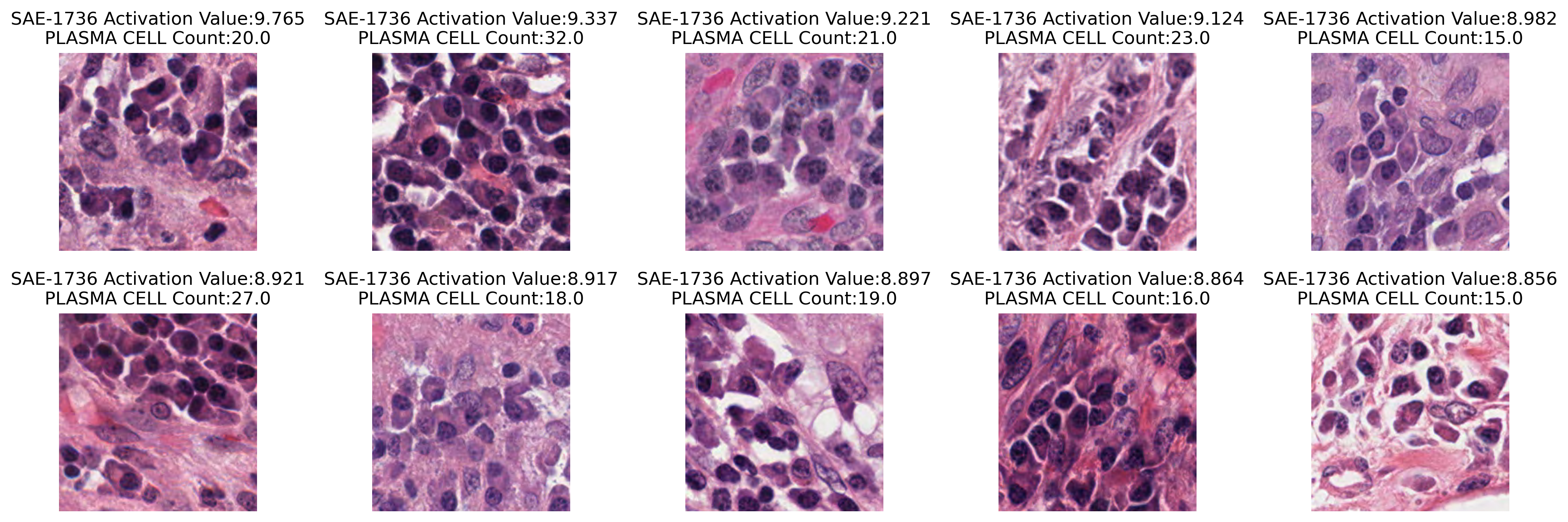}
    \caption{SAE-1736 captures plasma cell histology. Top-10 images with the highest SAE-1736 activation values and the corresponding plasma cell counts are shown.}
    \label{fig:Top_SAE-1736_images}
\end{figure*}

\begin{figure*}[h!]
\begin{center}
\centerline{\includegraphics[width=1.5\columnwidth]{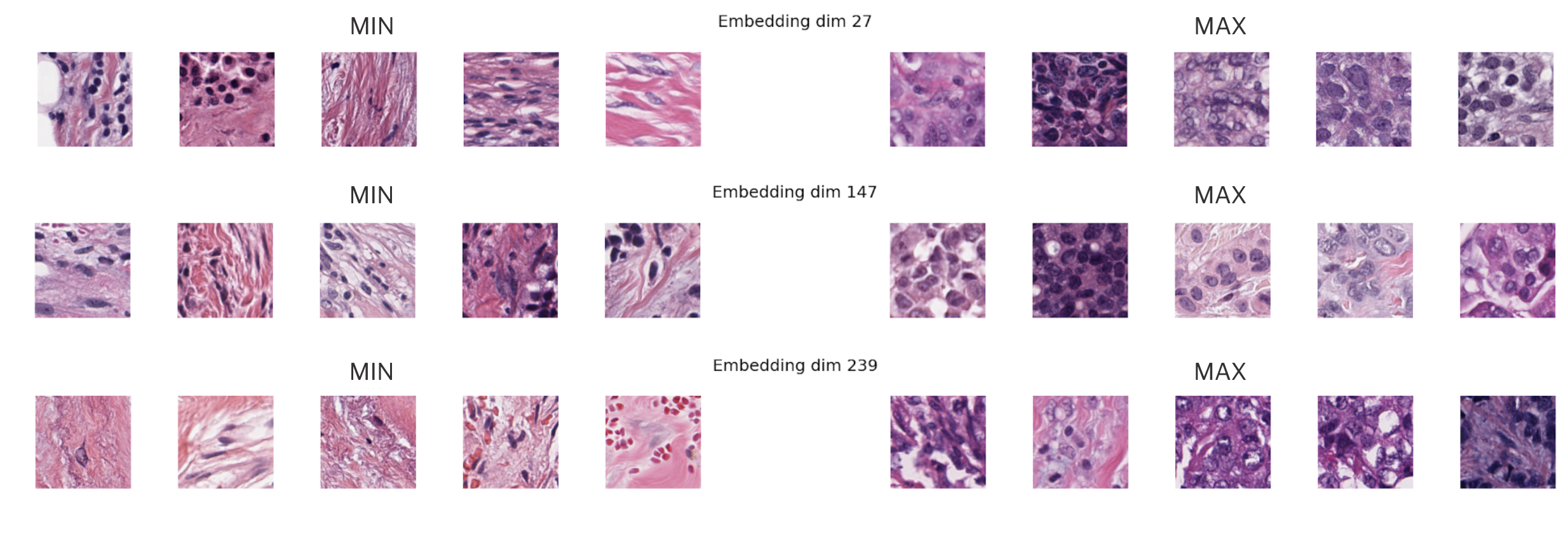}}
\caption{Visualization of features activating each embedding dimension. In each dimension, 5 example patches in the lowest 5\% and highest 5\% respectively of that dimension's activation are visualized. Inspection of each these patches reveals that multiple atomic features vary within each embedding dimension, including background stain color, cell size, shapes or morphologies. Some dimensions correspond to complex concepts that are relevant to pathology.}
\label{embedding_viz}
\end{center}
\end{figure*}

\begin{figure*}
    \centering
    \includegraphics[width=0.8\textwidth]{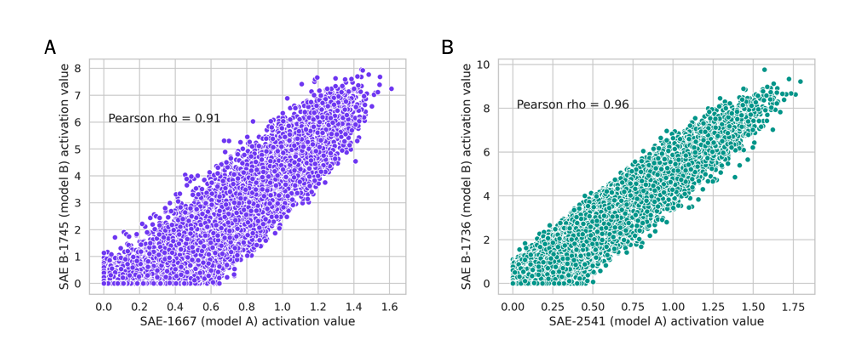}
    \caption{A) Anthracotic macrophage SAE feature comparison between model A and B. B) Plasma cell SAE feature comparison between model A and B. The high correlation values demonstrate that models trained on different datasets are able to uncover SAE dimensions that capture the same histological concepts.}
    \label{fig: SAE universality}
\end{figure*}

\end{document}